\documentclass[12 pt]{article} 
\usepackage{times,bm}
\usepackage[affil-it,]{authblk}
\usepackage[pagebackref=false,colorlinks=true,linkcolor=blue,citecolor=blue,urlcolor=blue]{hyperref}
\usepackage{geometry}
\usepackage{graphicx}
\usepackage{cite}
\usepackage{amsmath}
\usepackage{amsfonts}
\usepackage{amssymb}
\usepackage{color}
\usepackage{float}
\usepackage{multirow,multicol}
\usepackage{tabulary}
\usepackage{rotating}
\usepackage[flushleft]{threeparttable}
\usepackage{pgfplots,subfigure}
\usepackage{xcolor}
\usepackage{booktabs}
\numberwithin{equation}{section}
\usepackage{tikz}
\usepackage{ulem}
\usepackage{hyperref}
\usepackage{nameref}
\usepackage{comment}

\usepgflibrary{arrows}
\usetikzlibrary{shapes.callouts}
\tikzset{  
	level/.style   = { thick, },
	connect/.style = { dotted, red   },
	notice/.style  = { draw, rectangle callout, callout relative pointer={#1} },
	label/.style   = { text width=2cm }
}

\usepackage{letltxmacro}
\makeatletter
\let\oldr@@t\r@@t
\def\r@@t#1#2{%
	\setbox0=\hbox{$\oldr@@t#1{#2\,}$}\dimen0=\ht0
	\advance\dimen0-0.2\ht0
	\setbox2=\hbox{\vrule height\ht0 depth -\dimen0}%
	{\box0\lower0.4pt\box2}}
\LetLtxMacro{\oldsqrt}{\sqrt}
\renewcommand*{\sqrt}[2][\ ]{\oldsqrt[#1]{#2}}
\makeatother
\begin{document}

\newcommand{{\ri}}{{\rm{i}}}
\newcommand{{\Psibar}}{{\bar{\Psi}}}

\title{
		Thermodynamics and evaporation of a modified Schwarzschild black hole in
	a non--commutative gauge theory
	\\
}
\author{\large \textit {A. A. Araújo Filho  }$^{\ 1}$\footnote{E-mail: dilto@fisica.ufc.br }~,~ \textit {S. Zare}$^{\ 2}$\footnote{E-mail: soroushzrg@gmail.com}~,~ \textit {P. J. Porfírio}$^{\ 3}$\footnote{E-mail: pporfirio@fisica.ufpb.br}~,~ \textit {J.  K\v{r}\'{i}\v{z}}$^{\ 4}$\footnote{E-mail: jan.kriz@uhk.cz}~ and~\textit {H. Hassanabadi}$^{\ 2,4}$\footnote{E-mail: hha1349@gmail.com}  \\

	\small \textit {$^{\ 1}$ Departamento de Física Teórica and IFIC, Centro Mixto Universidad de Valencia--CSIC. Universidad de Valencia, Burjassot-46100, Valencia, Spain}\\
	\small \textit {$^{\ 2}$Faculty of Physics, Shahrood University of Technology, Shahrood, Iran}\\
	\small \textit{ $^{\ 3}$Departamento de Física, Universidade Federal da Paraíba, Caixa Postal 5008, 58051-970, João Pessoa, Paraíba,  Brazil}\\
	\small \textit{ $^{\ 4}$Department   of   Physics,   University   of   Hradec   Kr\'{a}lov\'{e}, Rokitansk\'{e}ho   62,   500   03   Hradec   Kr\'{a}lov\'{e},   Czechia}\\
}

\date{}
\maketitle

\begin{abstract}	
	
In this work, we study the thermodynamic properties on a non--commutative background via gravitational gauge field potentials. This procedure is accomplished after contracting de Sitter (dS) group, $\mathrm{SO}(4,1)$, with the Poincarè group, $\mathrm{ISO}(3,1)$. Particularly, we focus on a static spherically symmetric black hole. In this manner, we calculate the modified Hawking temperature and the other deformed thermal state quantities, namely, entropy, heat capacity, Helmholtz free energy and pressure. Finally, we also investigate the black hole evaporation process in such a context.

\end{abstract}

\begin{small}	
\end{small}


\bigskip

\newpage
\section{Introduction}

The formalism of classical thermodynamics can effectively be employed to investigate the physical aspects of black holes based on semiclassical methods of general relativity. The most famous studies in this direction were proposed by Hawking \cite{hawking1973large,hawking1971gravitational,hawking1975particle} and Bekenstein \cite{bekenstein1981universal,bekenstein1973black} with the purpose of solving the notorious information paradox \cite{boughn2022modest}. Remarkably, being supported by quantum field theory in curved spacetimes, they found that the event horizon area $A$ of a black hole turned out to be mathematically related to its corresponding entropy in the context of asymptotically--flat spacetimes. \cite{birrell1984quantum,parker2009quantum,nariai1999new}.

With the knowledge of thermodynamics, there can certainly be possible to address the problem of thermal stability for a given system . From the crucial work of Hawking and Gibbons \cite{gibbons1977cosmological}, it is well--known that such an issue may be extended to the black holes in non--asymptotically flat spacetimes \cite{escamilla2016thermodynamics}. In that reference, it is seen that the thermodynamic information of de Sitter black holes shows a significant discrepancy if compared to that ones in the context of an asymptotically flat spacetimes. Also, the authors  verified that, for a particular reference frame, the latter ones have a perfect black body radiation spectra, having their temperature calculated through the surface gravity $\kappa$. More so, within de Sitter spacetime, a cosmological event horizon is present and the emission of particle modes can also occur for a very particular situation in which both surface gravities are equal. This configuration only happens when there exists a degenerate case of extreme solutions \cite{nariai1950some,escamilla2016thermodynamics}.

Such a formalism does not properly possess a bound for the precision with which measurements of distances are done. In addition, it is believed that the bounds should exist, being given by the Planck length. In this sense, one of the most common approaches to accomplish such a feature is via non--commutativity of spacetimes. In an overall context, the main motivation to study non--commutative geometry stems from string/M--theory \cite{szabo2006symmetry,szabo2003quantum,3}. Other important applications arise in the context of supersymmetric Yang--Mills theories within the superfield approach \cite{ferrari2003finiteness,ferrari2004superfield,ferrari2004towards}. Particularly, in the context of gravity, this technique is typically inserted by using the Seiberg--Witten map, which gauges an appropriate group \cite{chamseddine2001deforming}. In the non--commutative approach, many works have been made in the context of black holes \cite{nicolini2009noncommutative,lopez2006towards,modesto2010charged,mann2011cosmological,1,2}, their evaporating aspects 
and their thermodynamic properties \cite{NicoliniPLB2006,NozariCQG2008,myung2007thermodynamics,PerezJMP2009,banerjee2008noncommutative,sharif2011thermodynamics,nozari2006reissner,nozari2007thermodynamics}. Nevertheless, up to now, there is a gap in the literature concerning a concise thermal study of the Schwarzschild black hole coming from the deformed gravitational gauge potentials.

In this work, based on the notable works of Chaichian et al. \cite{1} and Zet et al. \cite{2}, we are motivated to investigate the thermodynamic properties and the evaporation process \cite{NicoliniPLB2006,NozariCQG2008,PerezJMP2009} of a Schwarzschild black hole arising from the deformed gravitational gauge potentials (or tetrad fields), namely, $\hat{e}^{a}_{\mu} (x,\Theta)$ \cite{Touati2022}. Such a formalism is achieved by contracting the non--commutative gauge de Sitter (dS) group, $\mathrm{SO}(4,1)$, with the Poincarè (inhomogeneous Lorentz) group, $\mathrm{ISO}(3,1)$, through the Seiberg–Witten (SW) map approach \cite{3,4,5}. After that, we calculate the modified Hawking temperature and the other deformed thermal state quantities: entropy, heat capacity, Helmholtz free energy and pressure. 
In addition, we investigate the process of black hole evaporation \cite{NicoliniPLB2006,NozariCQG2008} within this context as well.




\section{Deformed gravitational gauge potentials}
In this section, we shall provide the main tools to address the non--commutative gauge theory of gravity. As it is said before, the gauge group corresponds to the dS one, $\mathrm{SO}(4,1)$. Let us briefly describe the  $\mathrm{SO}(4,1)$ gauge theory on a commutative $(3+1)$--dimensional Minkowski spacetime whose metric in spherical coordinates is given by
\begin{equation}
	\mathrm{d}s^{2}=\mathrm{d}r^{2}+r^{2}\mathrm{d}\Omega^{2}_{2}-c^{2}\mathrm{d}t^{2},
\end{equation}
with $\mathrm{d}\Omega^{2}_{2} = \left(\mathrm{d}\theta^{2}+\mathrm{sin}^{2}\theta\mathrm{d}\varphi^{2}\right)$. It is worth mentioning that the $\mathrm{SO}(4,1)$ group is a ten--dimensional group whose infinitesimal generators are indicated by $\mathcal{M}_{\mathcal{A}\mathcal{B}}=-\mathcal{M}_{\mathcal{B}\mathcal{A}}$, in which $\mathcal{A} = a, 5$ and $\mathcal{B} = b, 5$, where $a, b = 1, 2, 3, 0$. The generators $\mathcal{M}_{\mathcal{A}\mathcal{B}}$ may properly be distinguished from rotations $\mathcal{M}_{ab}=-\mathcal{M}_{ba}$ and translations $\mathcal{P}_{a} = \mathcal{M}_{a5}$.
The non--deformed gauge potentials, $\omega^{\mathcal{A}\mathcal{B}}_{\mu}(x) =-\omega^{\mathcal{B}\mathcal{A}}_{\mu}(x)$,  are distinguished from the spin connection, $\omega^{ab}_{\mu}(x) =-\omega^{ba}_{\mu}(x)$, and from the tetrad fields, $e^{a}_{\mu}(x)$, so that $\hat{\omega}^{a5}_{\mu}(x)=\mathcal{K} \hat{e}^{a}_{\mu}(x)$, in which $\mathcal{K}$ is a contraction parameter. 

Moreover, there is another gauge field via $\hat{\omega}^{55}_{\mu}(x)=\mathcal{K} \hat{\phi}_{\mu}(x,\Theta)$, which suffices to eliminate $\hat{\phi}_{\mu}(x,\Theta)$ by considering the limit $\mathcal{K}\rightarrow0$.
In other words, we deal with the Poincar\`{e} gauge group $\mathrm{ISO}(3,1)$, if we regard such a limit $\mathcal{K}\rightarrow0$ \cite{1,2}. The field strength related to  $\omega^{\mathcal{A}\mathcal{B}}_{\mu}(x)$ reads as follows 
\begin{equation}
	F^{\mathcal{A}\mathcal{B}}_{\mu} = \partial_{\mu}\omega^{\mathcal{A}\mathcal{B}}_{\nu} - \partial_{\nu}\omega^{\mathcal{A}\mathcal{B}}_{\mu} + \left(\omega^{\mathcal{A}\mathcal{C}}_{\mu}\omega^{\mathcal{D}\mathcal{B}}_{\nu}-\omega^{\mathcal{A}\mathcal{C}}_{\nu}\omega^{\mathcal{D}\mathcal{B}}_{\mu}\right)\eta_{\mathcal{C}\mathcal{D}}
\end{equation}
where $\mu, \nu = 1, 2, 3, 0$, and 
$\eta_{\mathcal{A}\mathcal{B}}=\mathrm{diag}(+++-+)$. More so, we can write
\begin{subequations}
	\begin{align}
		&F^{a5}_{\mu\nu}=\mathcal{K}\left[\partial_{\mu}e^{a}_{\nu}-\partial_{\nu}e^{a}_{\mu}+\left(\omega^{ab}_{\mu}e^{a}_{\nu}-\omega^{ab}_{\nu}e^{c}_{\mu}\right)\eta_{bc}\right]=\mathcal{K}T^{a}_{\mu\nu},\label{torsion}\\
		&F^{ab}_{\mu\nu} = \partial_{\mu} \omega^{ab}_{\nu}-\partial_{\nu}\omega^{ab}_{\mu}+\left(\omega^{ac}_{\mu}\omega^{db}_{\nu}-\omega^{ac}_{\nu}\omega^{db}_{\mu}\right)\eta_{cd}+\mathcal{K}\left(e^{a}_{\mu}e^{b}_{\nu}-e^{a}_{\nu}e^{b}_{\mu}\right)=R^{ab}_{\mu\nu},
	\end{align}
\end{subequations}
in which $\eta_{ab}=\mathrm{diag}(+++-)$. Here, it is worthy to be mentioned that the Poincar\'{e} gauge group under consideration possesses geometric structures coming from the Riemann--Cartan spacetime, including curvature and torsion fields \cite{1,6}. The torsion tensor $T^{a}_{\mu\nu}\equiv F^{a5}_{\mu\nu}/\mathcal{K}$ and the curvature tensor $R^{ab}_{\mu\nu}\equiv F^{ab}_{\mu\nu}$ associated with Riemann--Cartan spacetimes are fundamentally described by $e^{a}_{\mu}(x)$ and $\omega^{ab}_{\mu}(x)$. Given Eq. \eqref{torsion}, in the absence of torsion fields, one can obtain the spin connections in terms of tetrad fields. 

Now, let us start presenting a possible scenario of spherically gauge fields of the SO(4,1) group \cite{1,6}:
\begin{equation}\label{tetrads1}
	e^{1}_{\mu} = \left(\frac{1}{\mathcal{A}}, 0,0,0\right), \quad e^{2}_{\mu} = \left(0, r,0,0\right), \quad e^{3}_{\mu} = \left(0,0,r\, \mathrm{sin}\theta,0\right), \quad e^{0}_{\mu} = \left(,0,0,0, \mathcal{A}\right),
\end{equation}
and 
\begin{equation}\label{omega}
	\begin{split}
		& \omega^{12}_{\mu} = \left(0, \mathcal{W},0,0\right), \quad 
		\omega^{13}_{\mu} = \left(0,0, \mathcal{Z}\, \mathrm{sin}\theta,0\right),  \quad \omega^{10}_{\mu} = \left(0,0,0,\mathcal{U}\right),\\
		& \omega ^{23}_{\mu} = \left(0,0,-\mathrm{cos}\theta, \mathcal{V}\right), \quad \omega^{20}_{\mu} = \omega^{30}_{\mu} = \left(0,0,0,0\right),
	\end{split}
\end{equation}
where $\mathcal{A}$, $\mathcal{U}$, $\mathcal{V}$, $\mathcal{W}$ and $\mathcal{Z}$ are defined as a function of three--dimensional radius only. Furthermore, the non--null components of the torsion tensor are given as follows \cite{2}
\begin{equation}\label{torsion2}
	\begin{split}
		&T^{0}_{01} = -\frac{\mathcal{A}\mathcal{A}'+U}{\mathcal{A}}, \qquad 
		T^{2}_{03} = r\, \mathcal{V} \mathrm{sin}\theta \,T^{2}_{12} = \frac{\mathcal{A}+\mathcal{W}}{\mathcal{A}},\\
		& T^{3}_{02} = -r\, \mathcal{V}, \,\,\,\qquad\qquad T^{3}_{13} = \frac{\left(\mathcal{A}+\mathcal{Z}\right)\mathrm{sin}\theta}{\mathcal{A}},
	\end{split}
\end{equation}
and curvature tensor is \cite{2}
\begin{equation}\label{curvature}
	\begin{split}
		&R^{01}_{01} = \mathcal{U}', \quad R^{23}_{01} = -\mathcal{V}', \quad R^{13}_{23} = \left(\mathcal{Z}-\mathcal{W}\right) \mathrm{cos} \theta,\quad R^{01}_{01} = -\mathcal{U}\mathcal{W}, \quad R^{13}_{01} = - \mathcal{V}\mathcal{W},\\
		& R^{03}_{03} = -\mathcal{U} \mathcal{Z} \mathrm{sin} \theta, \quad R^{12}_{03} = \mathcal{V} \mathcal{Z} \mathrm{sin}\theta \, R^{12}_{12} = \mathrm{W}',\quad R^{23}_{23} = \left(1-\mathcal{Z}\mathcal{W}\right)\mathrm{sin}\theta, \quad R^{13}_{13} = \mathcal{Z}' \mathrm{sin}\theta.
	\end{split}
\end{equation}
Prime symbols $\mathcal{A}'$, $\mathcal{U}'$, $\mathcal{V}'$, $\mathcal{W}'$, and $\mathcal{Z}'$ denote derivatives with respect to radial components. Taking into account Eq. \eqref{torsion2}, we propose the following constraints to guarantee the absence of the torsion field 
\begin{equation}
	\mathcal{V} = 0, \qquad \mathcal{U} = - \mathcal{A}\mathcal{A}', \qquad \mathcal{W} =  -\mathcal{A} = \mathcal{Z}.
\end{equation}
With respect to the field equation 
\begin{equation}\label{fieldequation}
	R^{a}_{\mu} - \frac{1}{2} R\, e^{a}_{\mu} = 0,
\end{equation}
written in terms of the tetrad fields $e^{a}_{\mu}(x)$ with $R^{a}_{\mu} = R^{ab}_{\mu\nu} e^{\nu}_{b}$ and $R = R^{ab}_{\mu\nu} e^{\mu}_{a} e^{\nu}_{b}$, we obtain the following solution
\begin{equation}
	\mathcal{A}(r) = \sqrt{1-\frac{\alpha}{r}},
\end{equation}
where $\alpha$ is supposed to be an arbitrary constant given by $\alpha=2GM/c^{2}$, where the speed of light, the black hole mass, and the gravitational constant are represented by $c$, $M$, and $G$, respectively.
To find the relevant deformed metric $\mathrm{d}s^{2}=\hat{g}_{\mu\nu}(x,\Theta)\mathrm{d}x^{\mu}\mathrm{d}x^{\nu}$, with $x^{\mu} = r,\theta,\varphi, c\,t$, related to the Schwarzschild solution $(3+1)$--dimensional non--commutative spacetime, we need to have the deformed tetrad fields $\hat{e}^{a}_{\mu}(x,\Theta)$ arising from the contraction of the non--commutative gauge dS group $\mathrm{SO}(4,1)$, with the Poincar\`{e}  group, $\mathrm{ISO}(3,1)$, based on the SW map approach \cite{3,4,5}.
The non--commutative spacetime structure may be constructed to satisfy 
\begin{equation}\label{NonCommSTcond1}
	\left[x^{\mu},x^{\nu}\right]=i\Theta^{\mu\nu},
\end{equation}
where parameters $\Theta^{\mu\nu}$ are regarded to be real constants in such a way that  $\Theta^{\mu\nu}=-\Theta^{\nu\mu}$. 
Therefore, the gravitational fields, i.e., $\hat{e}^{a}_{\mu}(x,\Theta)$, and $\hat{\omega}^{\mathcal{A}\mathcal{B}}_{\mu}(x,\Theta)$, subjected to a non--commutative spacetime structure, are written in the power series of the parameter $\Theta$ \cite{1,2,3,4}
\begin{subequations}
	\begin{align}
		&\hat{e}^{a}_{\mu}(x,\Theta) = e^{a}_{\mu}(x)-i \Theta^{\nu\rho}e^{a}_{\mu\nu\rho}(x)+\Theta^{\nu\rho}\Theta^{\lambda\tau}e^{a}_{\mu\nu\rho\lambda\tau}(x)\dots,\label{tetradnoncom}\\	
		&\hat{\omega}^{\mathcal{A}\mathcal{B}}_{\mu}(x,\Theta) = \omega^{\mathcal{A}\mathcal{B}}_{\mu}(x)-i \Theta^{\nu\rho} \omega^{\mathcal{A}\mathcal{B}}_{\mu\nu\rho}(x)+\Theta^{\nu\rho}\Theta^{\lambda\tau}\omega^{\mathcal{A}\mathcal{B}}_{\mu\nu\rho\lambda\tau}(x)\dots  \,\,. \label{omeganoncom}
	\end{align}
\end{subequations} It should be noted that $\hat{e}^{a}_{\mu}(x,\Theta)$, in Eq. \eqref{tetradnoncom}, is obtained by using an expanded form of the following non--commutative corrections to $\hat{\omega}^{\mathcal{A}\mathcal{B}}_{\mu}(x,\Theta)$, given by Eq. \eqref{omeganoncom}, up to the second order  
\begin{subequations}\label{noncommcorr}
	\begin{align}
		\omega^{\mathcal{A}\mathcal{B}}_{\mu\nu\rho} (x) &= \frac{1}{4} \left\{\omega_{\nu},\partial_{\rho}\omega_{\mu}+R_{\rho\mu}\right\}^{\mathcal{A}\mathcal{B}},\label{noncommcorr-tetrad}\\
		\nonumber\omega^{\mathcal{A}\mathcal{B}}_{\mu\nu\rho\lambda\tau} (x) &= \frac{1}{32}\left(-\left\{\omega_{\lambda},\partial_{\tau}\left\{\omega_{\nu},\partial_{\rho}\omega_{\mu}+R_{\rho\mu}\right\}\right\}+2\left\{\omega_{\lambda},\left\{R_{\tau\nu},R_{\mu\rho}\right\}\right\}\right.\\
		\nonumber&\left.-\left\{\omega_{\lambda},\left\{\omega_{\nu},D_{\rho}R_{\tau\mu}+\partial_{\rho}R_{\tau\mu}\right\}\right\}-\left\{\left\{\omega_{\nu},\partial_{\rho}\omega_{\lambda}+R_{\rho\lambda}\right\},\left(\partial_{\tau}\omega_{\mu}+R_{\tau\mu}\right)\right\}\right.\\
		&\left.+2\left[\partial_{\nu}\omega_{\lambda},\partial_{\rho}\left(\partial_{\tau}\omega_{\mu}+R_{\tau\mu}\right)\right]\right)^{\mathcal{A}\mathcal{B}},\label{noncommcorr-omega}
	\end{align}
\end{subequations}
written after applying the SW map. Eqs. \eqref{noncommcorr-tetrad} and \eqref{noncommcorr-omega} obey the following relations
\begin{equation}
	\left[\alpha,\beta\right]^{\mathcal{A}\mathcal{B}} =  \alpha^{\mathcal{A}\mathcal{C}}\beta^{\mathcal{B}}_{\mathcal{C}}-\beta^{\mathcal{A}\mathcal{C}}\alpha^{\mathcal{B}}_{\mathcal{C}},\qquad
	\left\{\alpha,\beta\right\}^{\mathcal{A}\mathcal{B}} = \alpha^{\mathcal{A}\mathcal{C}}\beta^{\mathcal{B}}_{\mathcal{C}}+\beta^{\mathcal{A}\mathcal{C}}\alpha^{\mathcal{B}}_{\mathcal{C}},
\end{equation}
and also
\begin{equation}\label{}
	D_{\mu}R^{\mathcal{A}\mathcal{B}}_{\rho\sigma} = \partial_{\mu}R^{\mathcal{A}\mathcal{B}}_{\rho\sigma} +\left(\omega^{\mathcal{A}\mathcal{C}}_{\mu}R^{\mathcal{D}\mathcal{B}}_{\rho\sigma}+\omega^{\mathcal{B}\mathcal{C}}_{\mu}R^{\mathcal{D}\mathcal{A}}_{\rho\sigma}\right)\eta_{\mathcal{C}\mathcal{D}}.
\end{equation}
It is important to note that we highlight some constraints related to $\hat{\omega}^{\mathcal{A}\mathcal{B}}_{\mu}(x,\Theta)$: 
\begin{equation}\label{CondDefOmega}
	\hat{\omega}^{\mathcal{A}\mathcal{B}\star}_{\mu}(x,\Theta) = -\hat{\omega}^{\mathcal{A}\mathcal{B}}_{\mu}(x,\Theta), \quad 
	\hat{\omega}^{\mathcal{A}\mathcal{B}}_{\mu}(x,\Theta) ^{r} \equiv \hat{\omega}^{\mathcal{A}\mathcal{B}}_{\mu}(x,-\Theta) = -\hat{\omega}^{\mathcal{B}\mathcal{A}}_{\mu}(x,\Theta),
\end{equation}
where the ${}^\star$ superscript is a symbol which refers to the complex conjugate. Moreover, the corresponding non--commutative corrections stemming from constraints \eqref{CondDefOmega} can be expressed as
\begin{equation}
	\omega^{\mathcal{A}\mathcal{B}}_{\mu} (x) = - \omega^{\mathcal{B}\mathcal{A}}_{\mu} (x), \quad \omega^{\mathcal{A}\mathcal{B}}_{\mu\nu\rho} (x) = \omega^{\mathcal{B}\mathcal{A}}_{\mu\nu\rho} (x), \quad \omega^{\mathcal{A}\mathcal{B}}_{\mu\nu\rho\lambda\tau} (x) = -\omega^{\mathcal{B}\mathcal{A}}_{\mu\nu\rho\lambda\tau} (x).
\end{equation}
We realize that Eq. \eqref{tetradnoncom} is provided by using Eqs. \eqref{noncommcorr-tetrad} and \eqref{noncommcorr-omega} with the absence of torsion field $T^{a}_{\mu\nu}$, and the limit $\mathcal{K}\rightarrow0$. For Eq. \eqref{tetradnoncom}, the complex conjugate of deformed tetrad fields can be written as 
\begin{equation}\label{ComConjDefTetrads}
	\hat{e}^{a\star}_{\mu}(x,\Theta) = e^{a}_{\mu}(x)+i \Theta^{\nu\rho}e^{a}_{\mu\nu\rho}(x)+\Theta^{\nu\rho}\Theta^{\lambda\tau}e^{a}_{\mu\nu\rho\lambda\tau}(x)\dots,
\end{equation}
where
\begin{equation}
	\begin{split}
		e^{a}_{\mu\nu\rho} &= \frac14	\left[\omega^{ac}_{\nu}\partial_{\rho} e^{d}_{\mu}+\left(\partial_{\rho}\omega^{ac}_{\mu}+R^{ac}_{\rho\mu}\right)e^{d}_{\nu}\right]\eta_{ad},
	\end{split}
\end{equation}
and
\begin{equation}
	\begin{split}
		e^{a}_{\mu\nu\rho\lambda\tau}(x) &= \frac{1}{32}\left[2\left\{R_{\tau\nu},R_{\mu\rho}\right\}^{ab}e^{c}_{\lambda} - \omega^{ab}_{\lambda}\left(D_{\rho}+\partial_{\rho}\right)R^{cd}_{\tau\mu}e^{m}_{\nu}\eta_{dm}\right.\\
		&\left.-\left\{\omega_{\nu},\left(D_{\rho}+\partial_{\rho}\right)R_{\tau\mu}\right\}^{ab}e^{c}_{\lambda}-\partial_{\tau}\left\{\omega_{\nu},\left(\partial_{\rho}\omega_{\mu}+R_{\rho\mu}\right)\right\}^{ab}e^{c}_{\lambda}\right.\\
		&\left.-\omega^{ab}_{\lambda}\partial_{\tau}\left(\omega^{cd}_{\nu}\partial_{\rho}e^{m}_{\mu}+\left(\partial_{\rho}\omega^{cd}_{\mu}+R^{cd}_{\rho\mu}\right)e^{m}_{\nu}\right)
		\eta_{dm}+2\partial_{\nu}\omega^{ab}_{\lambda}\partial_{\rho}\partial_{\tau}e^{c}_{\mu}\right.\\
		&\left. -2\partial_{\rho}\left(\partial_{\tau}\omega^{ab}_{\mu}+R^{ab}_{\tau\mu}\right)\partial_{\nu}e^{c}_{\lambda}-\left\{\omega_{\nu},\left(\partial_{\rho}\omega_{\lambda}+R_{\rho\lambda}\right)\right\}^{ab}\partial_{\tau}e^{c}_{\mu}\right.\\
		&\left.-\left(\partial_{\tau}\omega^{ab}_{\mu}+R^{ab}_{\tau\mu}\right)\left(\omega^{cd}_{\nu}\partial_{\rho}e^{m}_{\lambda}+\left(\partial_{\rho}\omega^{cd}_{\lambda}+R^{cd}_{\rho\lambda}\right)e^{m}_{\nu}\eta_{dm}\right)
		\right]\eta_{bc}.
	\end{split}
\end{equation}
Thus, a deformed metric tensor can be written as
\begin{equation}\label{DefMetTensor}
	\hat{g}_{\mu\nu}\left(x,\Theta\right) = \frac12 \eta_{ab}\left(\hat{e}^{a}_{\mu}(x,\Theta)\ast\hat{e}^{b\star}_{\nu}(x,\Theta)+\hat{e}^{b}_{\mu}(x,\Theta)\ast\hat{e}^{a\star}_{\nu}(x,\Theta)\right),
\end{equation}
where symbol $\ast$ denotes the well--known star product. From now on, we shall use the natural units $\hbar=c=G=1$ in order to accomplish our calculations.

\section{Deformed Schwarzschild black hole
	\label{sec2}}

After taking into account Eqs. \eqref{tetrads1} \eqref{omega}, \eqref{DefMetTensor}, \eqref{tetradnoncom} and \eqref{ComConjDefTetrads}, we are properly able to introduce non--null components of the deformed metric tensor $\hat{g}_{\mu\nu} (x,\Theta)$, up to the second order of parameter $\Theta$  \cite{1,PerezJMP2009}. With this, we may address to the Schwarzschild solution of a non--commutative $(3+1)$--dimensional spherically symmetric spacetime  
\begin{subequations}
	\begin{align}
		\hat{g}_{11}(r,\Theta)&=\frac{1}{\mathcal{A}^{2}(r)}+\frac14\frac{\mathcal{A}''(r)}{\mathcal{A}(r)} \Theta^{2}+O\left(\Theta^{4}\right) \nonumber \\
		&\,\,\equiv \frac{1}{1-\frac{2M}{r}}+\left[\frac{3M^{2}-2 M r}{4 r^{2}\left(r-2M\right)^{2}}\right]\Theta^{2}+O\left(\Theta^{4}\right),\\
		\hat{g}_{22}(r,\Theta)&=r^{2}+\frac{1}{16}\left[\mathcal{A}^{2}(r)+11r\mathcal{A}(r)\mathcal{A}'(r)+16r^{2}\mathcal{A}'^{2}(r)\right.\nonumber\\
		&\left.\qquad\qquad\quad+12r^{2}\mathcal{A}(r)\mathcal{A}''(r)\right]\Theta^{2}+O\left(\Theta^{4}\right) \nonumber \\
		&\,\,\equiv r^{2}-\left[\frac{34M^{2}-17 M r+r^{2}}{16 r\left(2M-r\right)}\right]\Theta^{2}+O\left(\Theta^{4}\right),\\
		\hat{g}_{33}(r,\Theta)&=r^{2}\mathrm{sin}^{2}\theta+\frac{1}{16}\left[4\left(2r\mathcal{A}(r)\mathcal{A}'(r)-r\frac{\mathcal{A}'(r)}{\mathcal{A}(r)}+r^{2}\mathcal{A}(r)\mathcal{A}''(r)\right.\right. \nonumber\\
		&\left.\left.\qquad\qquad\quad+2r^{2}\mathcal{A}'^{2}(r)\right)\mathrm{sin}^{2}\theta +\mathrm{cos}^{2}\theta\right]\Theta^{2}+O\left(\Theta^{4}\right)
		\nonumber \\
		&\,\,\equiv r^{2}\mathrm{sin}^{2}\theta-\left[\frac{\left(r-2M\right)r\mathrm{cos}^{2}\theta+4M\left(M-r\right)\mathrm{sin}^{2}\theta}{16 r\left(2M-r\right)}\right]\Theta^{2}+O\left(\Theta^{4}\right),\\
		\hat{g}_{00}(r,\Theta)&=-\mathcal{A}^{2}(r)-\frac14\left[2r\mathcal{A}(r)\mathcal{A}'^{3}(r)+r\mathcal{A}^{3}(r)\mathcal{A}'''(r)+\mathcal{A}^{3}(r)\mathcal{A}''(r)\right.\nonumber\\
		&\left.\qquad\qquad\quad+2\mathcal{A}^{2}(r)\mathcal{A}'^{2}(r)+5r\mathcal{A}^{2}(r)\mathcal{A}'(r)\mathcal{A}''(r)\right] \Theta^{2}+O\left(\Theta^{4}\right) \nonumber\\
		&\equiv -1+\frac{2M}{r}+\left[\frac{11M^{2}-4 M r}{4 r^{4}}\right]\Theta^{2}+O\left(\Theta^{4}\right) .\label{gtt}
	\end{align}
\end{subequations}
As it can be seen, the deformed metric tensor $\hat{g}_{\mu\nu} (x,\Theta)$ is diagonal.
It should be noted that the non--commutative structure of spacetime is specified by
Eq. \eqref{NonCommSTcond1}. 
In this scenario, the deformed diagonal metric $\hat{g}_{\mu\nu} (x,\Theta)$ is provided by considering the non--null components of the $\Theta^{\mu\nu}$ tensor
\begin{equation}\label{NonCommST-Tensor}
	\Theta^{12}=\Theta=-\Theta^{21},
\end{equation}
where $\Theta$ specifies  the non--commutativity of the spacetime coordinates. 

\begin{figure}[H]
		\begin{center}
			\includegraphics[scale=0.5]{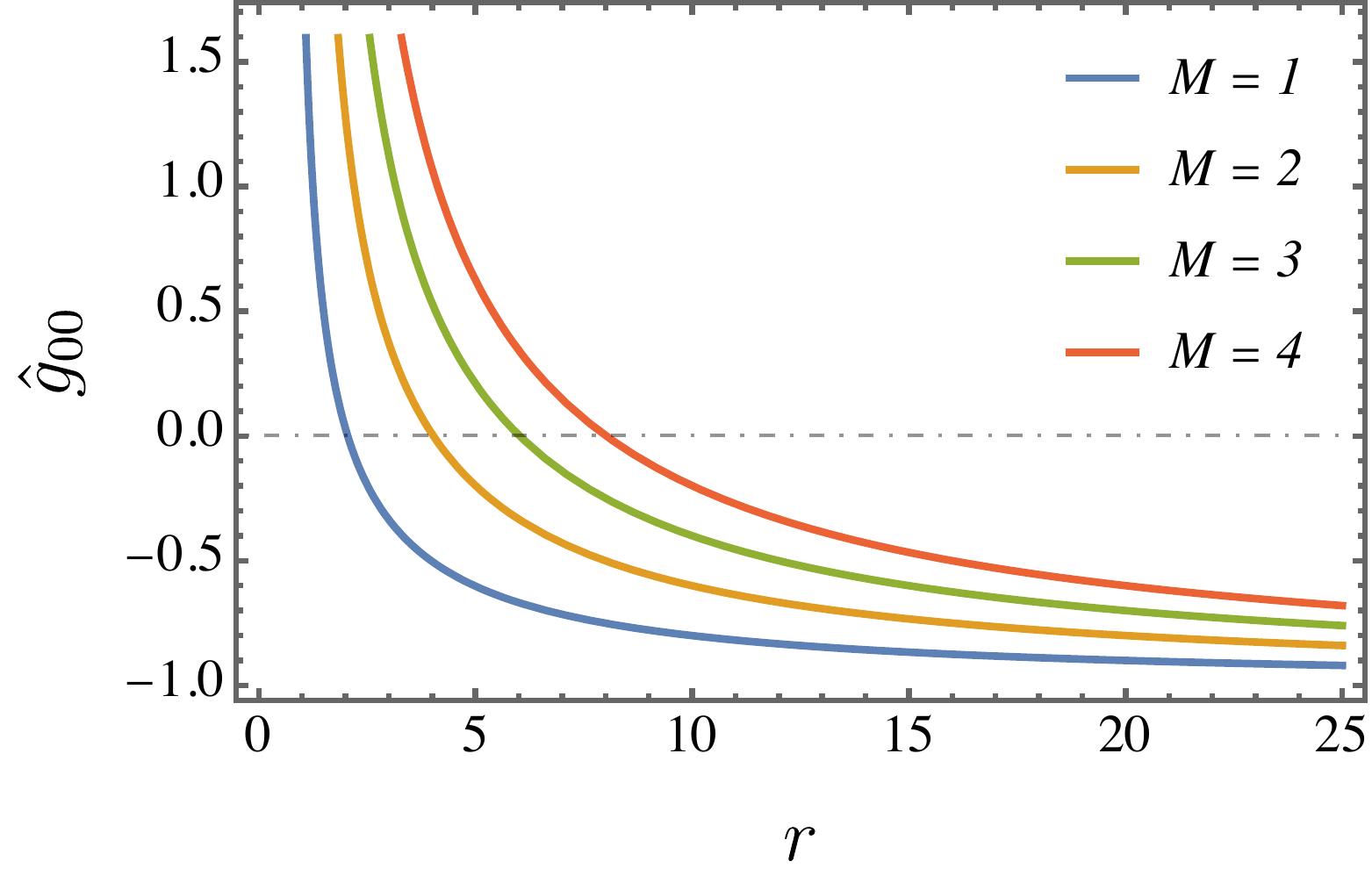}
			\caption{
				$\hat{g}_{00}$ as a function of radius $r$ for three different values of mass $M$ with $\Theta = 0.8$.
				\label{Fig1}}
		\end{center}
	\end{figure}

	In Figure \ref{Fig1}, we plot $\hat{g}_{00}$ versus $r$ for different $M$. The figure shows that there is only one horizon for each particular value of $M$. Besides, the figure indicates that when $M$ runs, the amount of the horizon increases. For all cases, we take into account a specific value for parameter $\Theta$, namely, $\Theta=0.8$.

If we consider the non--null component exhibited in  Eq. \eqref{gtt} being equal to zero, $\hat{g}_{00}(r,\Theta)|_{r=r_{s\Theta}}=0$, we can acquire the event horizon of the deformed Schwarzschild radius as 
	\begin{equation}\label{modradius}
		r_{s\Theta}=2M+\frac{3\Theta^{2}}{32M},
	\end{equation}
where $2M$ is the usual Schwarzschild radius $r_{s}$. 
	The mass of the Schwarzschild black hole in the framework of the non--commutative gauge theory of gravitation, i.e., a special case of non--commutative gravity, is obtained up to the second order of parameter $\Theta$ by $M_{\Theta}=r_{s\Theta}/2$. Naturally, we recover the usual mass of the Schwarzschild black hole, if we consider the limit where $\Theta\rightarrow0$.

For our purpose, we should define its corresponding horizon area as
	\begin{equation}
		A_{\Theta}=\int_{0}^{2\pi}\int_{0}^{\pi}\sqrt{\hat{g}_{22}\hat{g}_{33}}\mathrm{d}\theta\mathrm{d}\varphi ,
	\end{equation}
	so that we can obtain the following expression
\begin{equation}\label{HorizonArear}
	\begin{split}
A_{\Theta} &=2\pi \left[\int_{0}^{\pi}\left(r_{s\Theta}^{2}-\frac{16}{64r_{s\Theta}}\Theta^{2}\right)\mathrm{sin}\theta\mathrm{d}\theta+\int_{0}^{\pi}\frac{\Theta^{2}}{32}\mathrm{csc}\theta\mathrm{d}\theta\right]\\ 
&=4\pi r^{2}_{s}+\frac{5\pi}{16}\,\Theta^{2},	
	\end{split}
\end{equation}
where we have used the residual amount for calculating the third term.

In the literature, there exist some methods to obtain the black hole entropy \cite{PerezJMP2009,7,8}. 
Here, it worth mentioning that the thermodynamic state quantities were also calculated in other different cosmological scenarios \cite{araujo2021bouncing,filho2022thermodynamics,araujo2022particles,araujo2021higher,aa2021lorentz,araujo2022thermal,reis2021thermal,9}.

	Particularly, we can define the entropy of the black hole to our case by setting $S_{\Theta}=A_{\Theta}/4$. Then, it can be expressed in the terms of horizon area 
	\begin{equation}\label{nonCommEntropy1}
		S_{\Theta} = \pi r^{2}_{s}+\frac{5\pi}{64}\,\Theta^{2}.
	\end{equation}

Next, for the sake of providing another application to this scenario, we exhibit the deformed Schwarzschild black hole temperature \cite{NicoliniPLB2006,NozariCQG2008} within the semiclassical framework:

\begin{equation}\label{TempBH}
	T_{\Theta}=\frac{\partial M_{\Theta}}{\partial S_{\Theta}}=\frac{\partial M_{\Theta}/\partial M}{\partial S_{\Theta}/\partial M}.
\end{equation}
It should be noted that Hawking showed the relation between the temperature and its mass, having the following form: $T=\hbar c^{3}\kappa/2\pi k_{B}G$, where $\kappa$ is the surface gravity and $k_{B}$ is the Boltzmann constant. According to the Hawking's formula, if the temperature goes to infinity, its associated mass will tend to zero. In other words, it means that the black hole should evaporate. Nevertheless, a precise answer to this issue may only be possible through a final theory of quantum gravity. However, the non--commutative gauge theory of gravity offers a robust theoretical framework in order to overcome this dilemma. Now, considering its corrections up to the second order in $\Theta$, Eq. \eqref{TempBH} can be calculated as

	\begin{equation}\label{TempBH2}
		T_{\Theta}=\frac{1}{8\pi M}-\frac{3}{512 \pi M^{3}}\Theta^{2}.
	\end{equation}
We can see that the corrected temperature in Eq. \eqref{TempBH2} depends on parameter $\Theta$ and the usual Schwarzschild black hole mass $M$. 
Furthermore, we should note that $ \lim_{\Theta \to 0} T_{\Theta} = T$.

\begin{figure}[H]
		\begin{center}
			\includegraphics[scale=0.44]{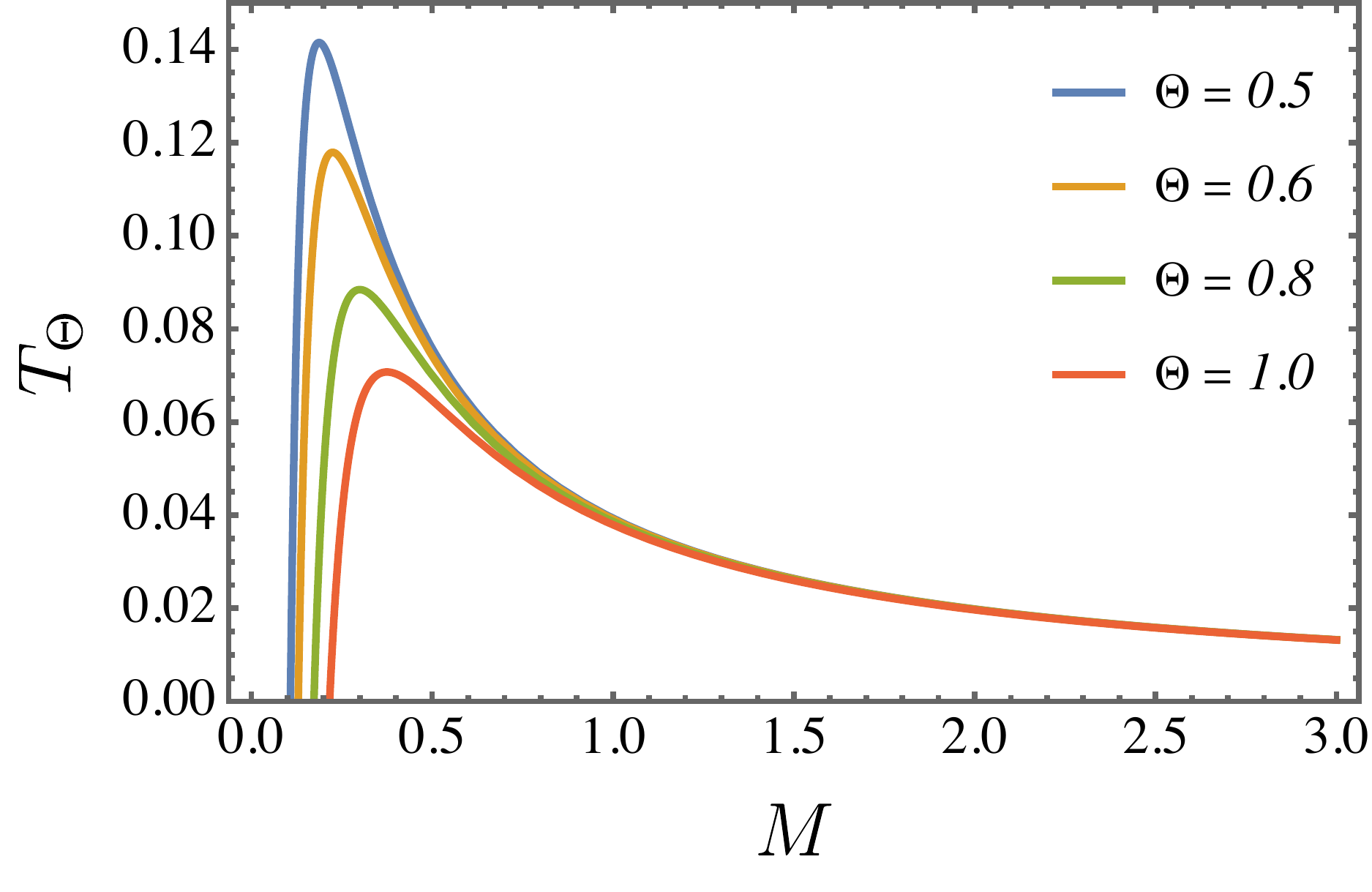}
   \includegraphics[scale=0.44]{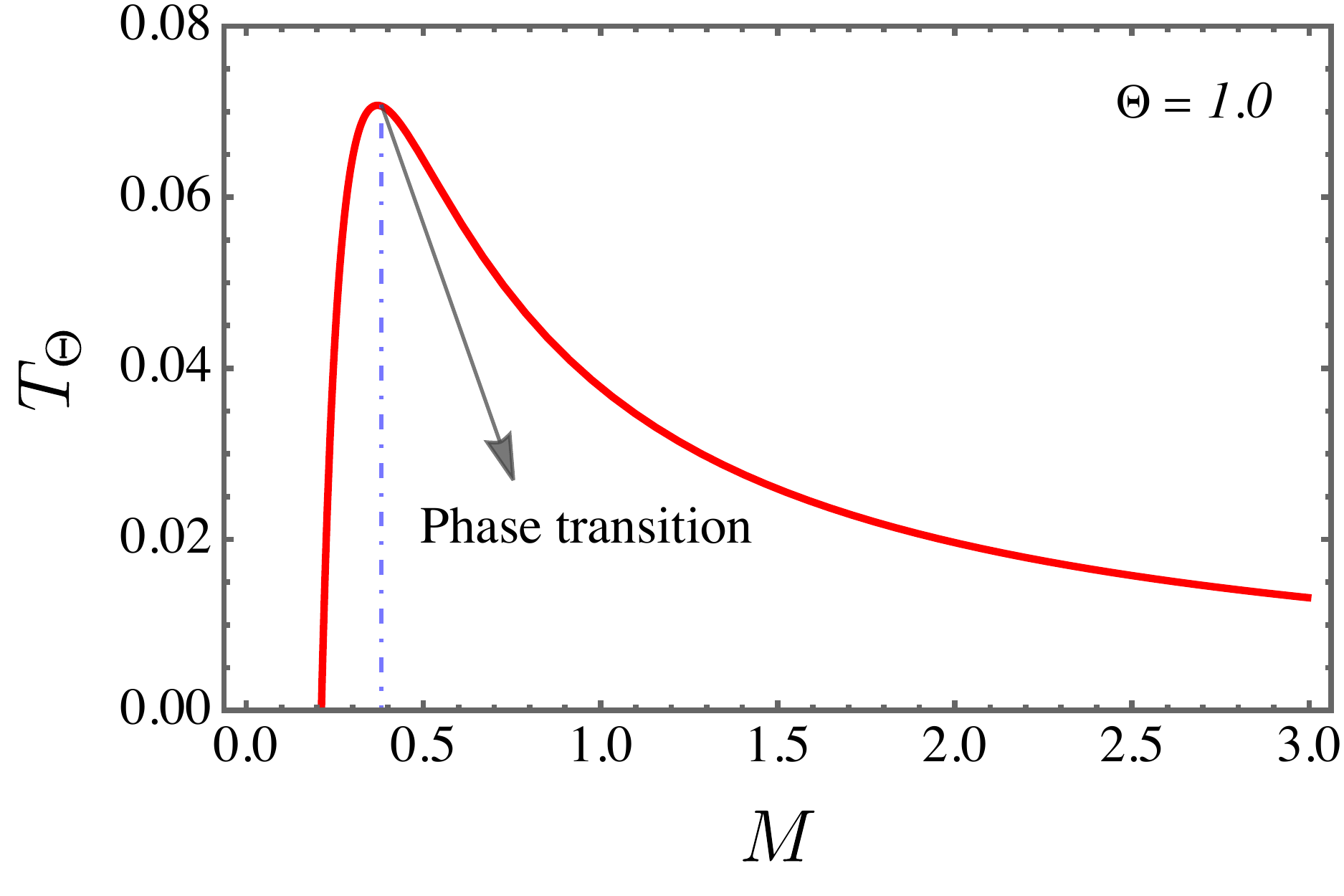}
			\caption{
				$T_{\Theta}$ as a function of mass $M$ for different values of $\Theta$ (on the left) and for a specific value of $\Theta$, namely, $\Theta=1$ (on the right). With this latter value, the phase transition occurs at $(0.375,0.071)$.
				 \label{Fig2}}
		\end{center}
	\end{figure}

In Figure \ref{Fig2}, we have plotted $T_{\Theta}$ versus $M$, which shows a minimal non--zero mass, $M_{min}=\sqrt{3}\Theta/8=M_{rem}$, with $M_{rem}$ being categorized as the remnant mass. It depends clearly on the amount of non--commutative parameter $\Theta$. Such a mass is the minimum mass value to which the black hole shrinks. Also, there is a maximum value for $T_{\Theta}$. This feature shows a phase transition, which entails that its corresponding heat capacity is zero at this point.

\begin{figure}[H]
\begin{center}
\includegraphics[scale=0.44]{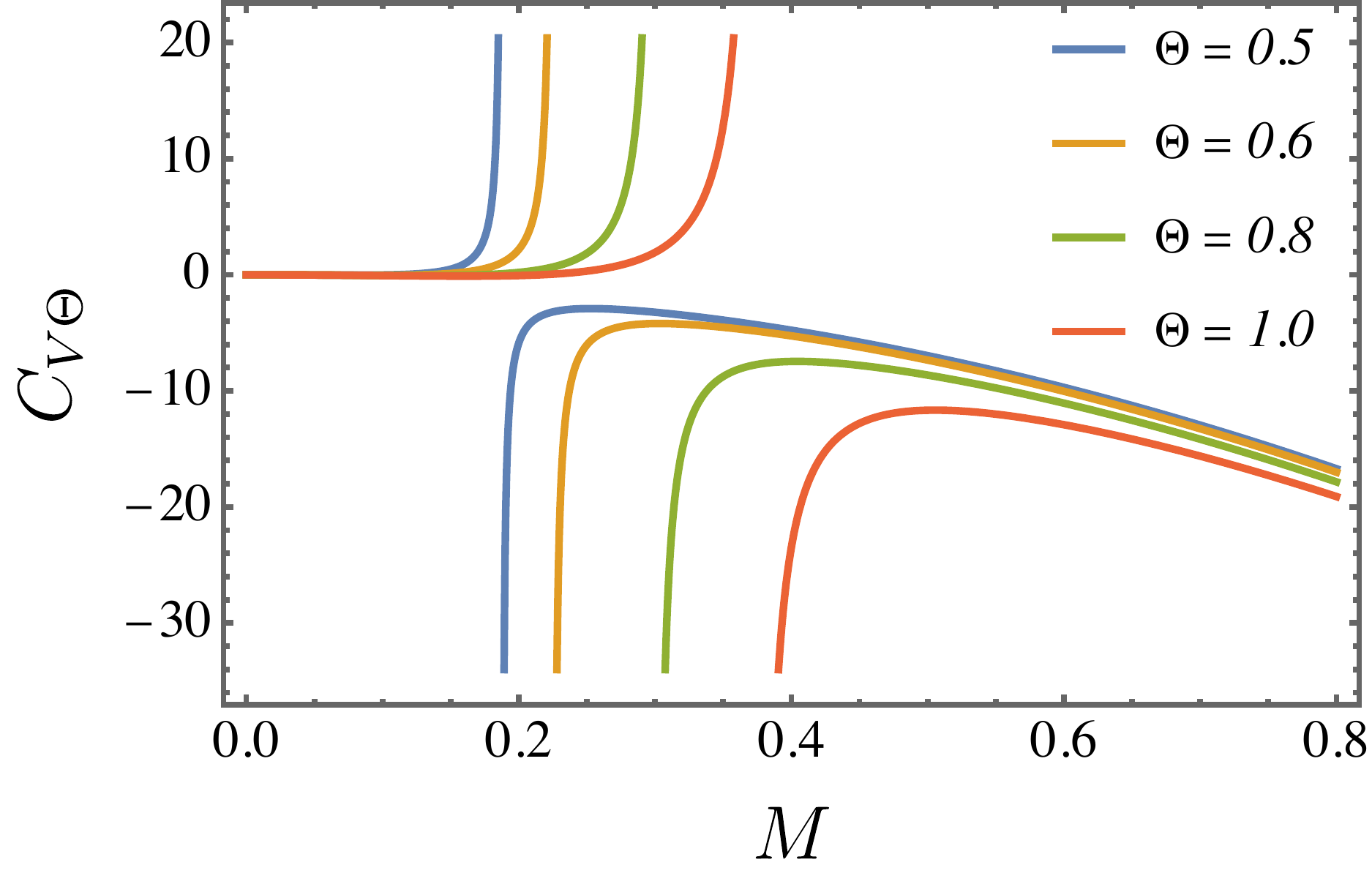}
\includegraphics[scale=0.44]{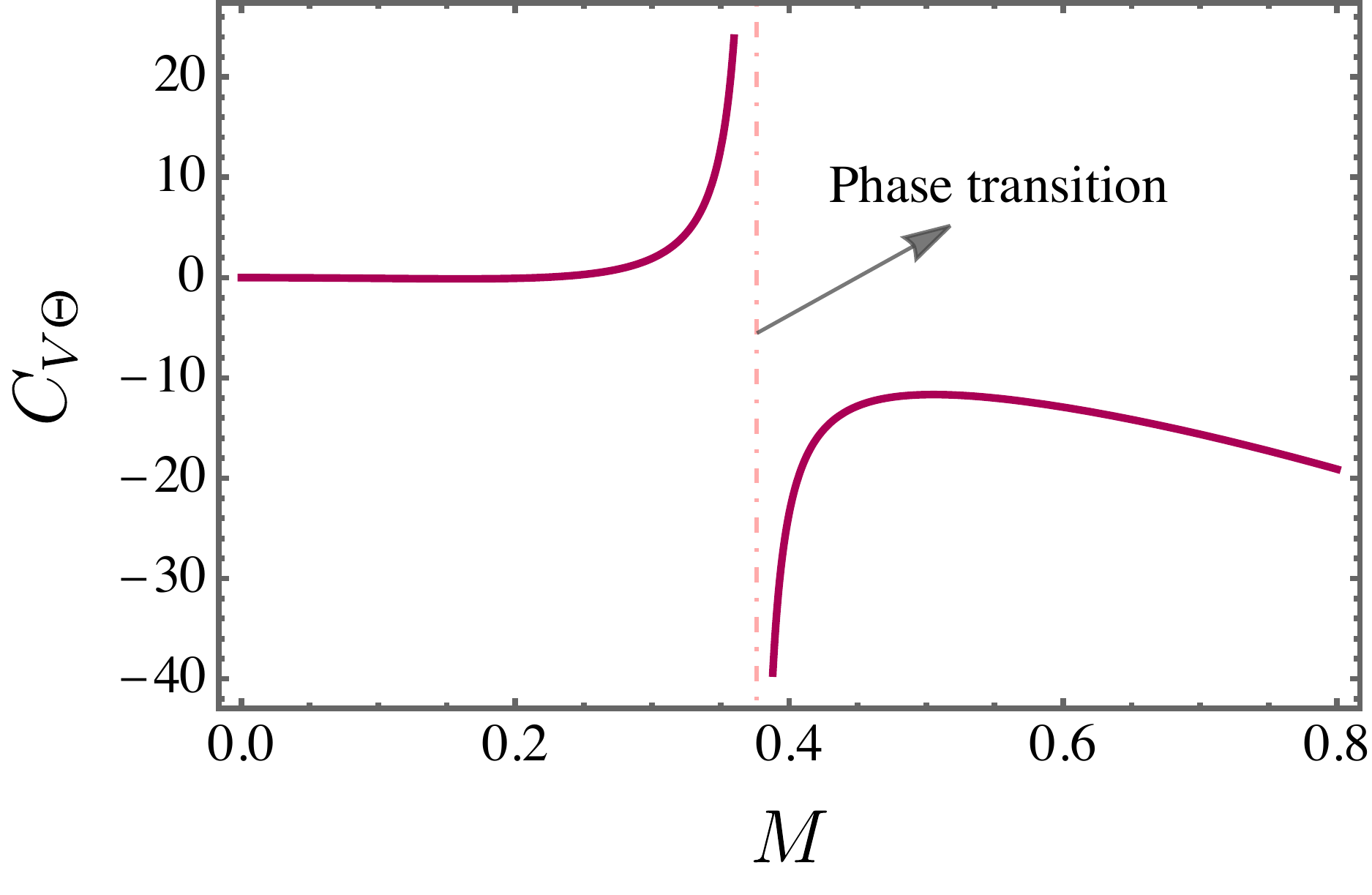}
\caption{
$C_{V\Theta}$ as a function of mass $M$ for different values of $\Theta$ (on the left) and for a specific value of $\Theta$, namely, $\Theta=1$ (on the right). The divergence occurs in the vicinity of the point $(0.375,0.071)$, which indicates the incidence of a phase transition.\label{Fig3}}
\end{center}
\end{figure}

Furthermore, the heat capacity can be obtained by
\begin{equation}
	C_{V\Theta} = T_{\Theta}\frac{\partial S_{\Theta}}{\partial T_{\Theta}} = T_{\Theta} \frac{\partial S_{\Theta}/\partial M}{\partial T_{\Theta}/\partial M},
\end{equation}
 which leads to the following expressing:
\begin{equation}
    C_{V\Theta} = - \frac{8M^{2}\pi(64 M^{2}-3\Theta^{2})}{64M^{2}-9\Theta^{2}}.
\end{equation}
In Figure, \ref{Fig3} we have plotted the deformed heat capacity versus $M$.
It is worth emphasizing that this thermodynamic function addresses information about the stability of the black hole. Essentially,
for the region where the corresponding heat capacity is positive, there exists a local stability. On the other hand, in the region which possesses negative values, the opposite behavior gives rise to: the appearance of an unstable behavior \cite{Jawad2022NPB}. However, when the heat capacity is zero, it means that there will be no further change of mass through its temperature. In this case, the radiation process ends and the collapse of the black hole terminates, i.e., the mass remains unchanged. This mass is called the remnant mass, $M_{rem}$, as we pointed out previously \cite{BanerjeePLB2010}.
In addition, in the regions where such thermodynamic function has a discontinuity, there exists an appearance of a phase transition. According to Figure \ref{Fig3}, for $M=0.375$, we see such a feature. For a better comprehension of the reader, we will label this crucial point as $M_{pht}$ -- the phase transition mass point. Considering the region with the mass values are grater than $M_{pht}$, we expect to encounter instability while otherwise, we observe a stable behavior. Actually, if we take $C_{V\Theta}$ equal to zero, we reach $M_{rem}=\frac{\sqrt{3}\Theta}{8}$, which becomes $0.217$ for $\Theta=1$. The existence of a remnant mass $M_{rem}$ and its direct dependence on $\Theta$ indicates that the black hole totally will not evaporate. As a consequence, part of its information will be maintained in the form of a stable black hole remnant mass.

Furthermore, we can write the first law of thermodynamics as
\begin{equation}\label{FristLawTherm1}
	\mathrm{d}U = T \mathrm{d}S+\sum_{i=1}^{} \mathcal{X}_{i}\mathrm{d}y_{i} -\mathrm{d}(PV),
\end{equation}
in which the internal energy is denoted by  $U(y_{i})$, $\mathcal{X}_{i}$ mimics the $V$ volume, i.e., being a generic intensive quantity, the angular velocity $\Omega$, the electrostatic potential $\Phi$, and $y_{i}$ is the conjugate extensive parameter of the system. In our case, the horizon area and the volume of the black hole are geometrically associated; for instance, such an area of the deformed Schwarzschild black hole is shown in Eq. \eqref{HorizonArear} and the corresponding volume is given by
$V_{\Theta}=\frac{4}{3}\pi r^{3}_{s\Theta}$.

With these explanations, the non--commutative Helmholtz free energy $\mathcal{F}_{\Theta}$ can be obtained by $\mathcal{F}_{\Theta} = M_{\Theta} - T_{\Theta}S_{\Theta}$. In this sense, with respect to Eqs. \eqref{modradius}, \eqref{TempBH2} and \eqref{nonCommEntropy1}, it reads
\begin{equation}\label{DefFreeEnergy1}
\begin{split}
\mathcal{F}_{\Theta} &=\frac{M}{2} +\frac{31}{512M}\, \Theta^{2},
\end{split}
\end{equation}
and we keep the simplified form of Eq. \eqref{DefFreeEnergy1} up to the second order of $\Theta$.

\begin{figure}[H]
\begin{center}
\includegraphics[scale=0.5]{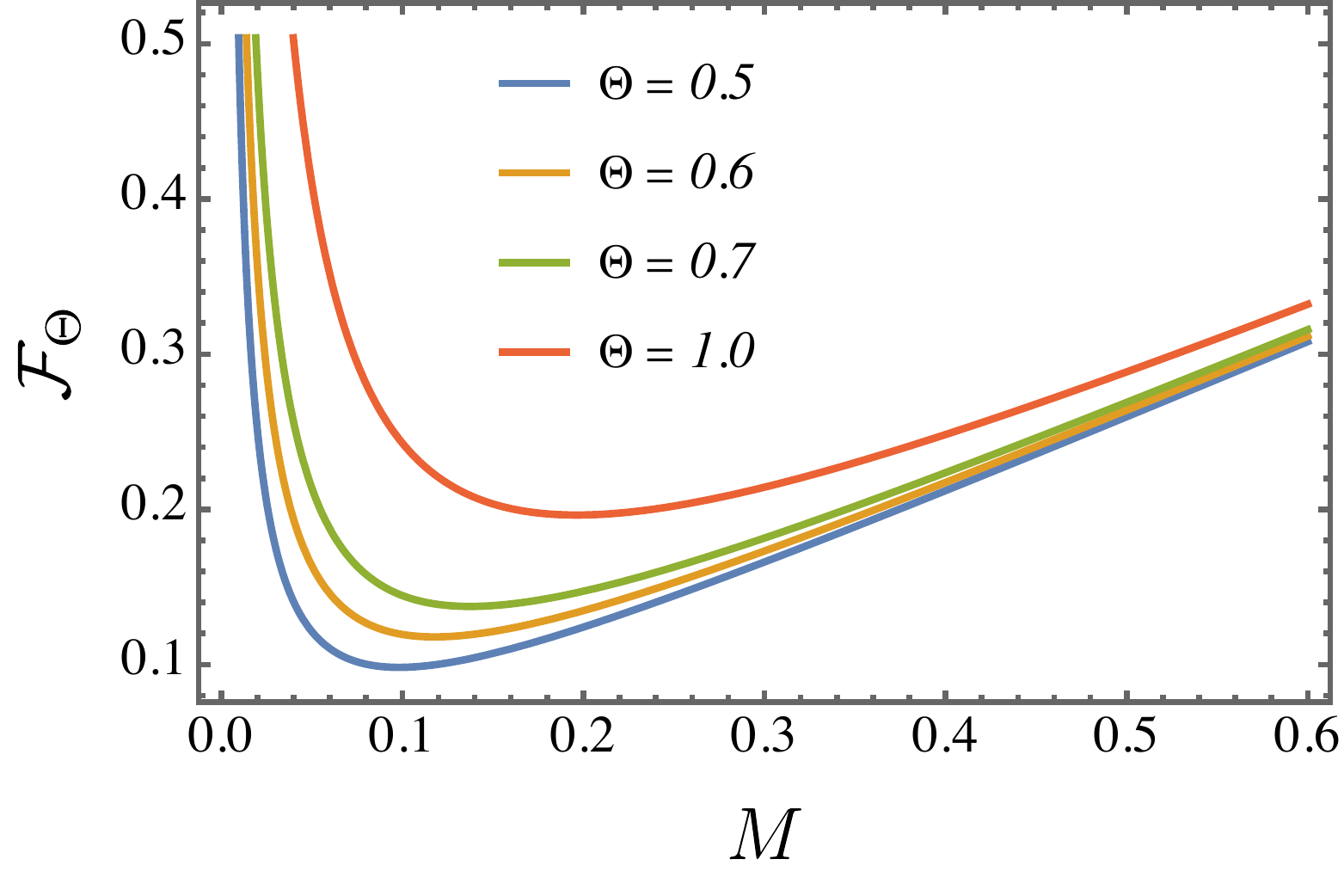}
\includegraphics[scale=0.5]{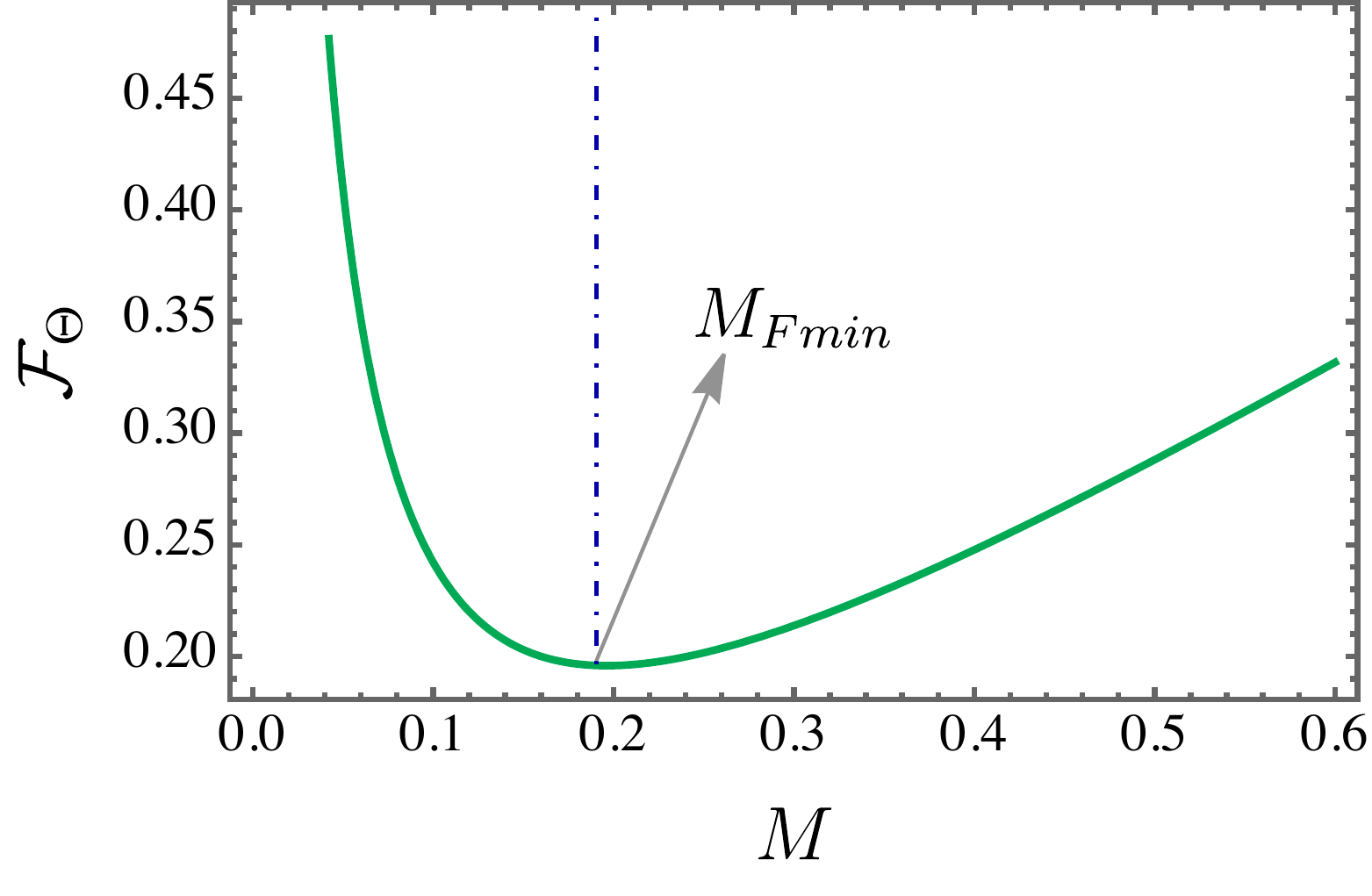}
\caption{
$\mathcal{F}_{\Theta}$ as a function of mass $M$ for three different values of $\Theta$ (on the left). On the other hand, it is also shown such thermodynamic function for a specific value of $\Theta$, i.e., $\Theta = 1$, (on the right). 
\label{Fig4}}
\end{center}
\end{figure}

In Figure \ref{Fig4}, we plot the non--commutative Helmholtz free energy versus $M$. Notice that this thermodynamic function has a minimum at $M_{Fmin} = (0.190, 0.197)$. Clearly, one can see that the values of $\mathcal{F}_{\Theta}$ decrease in the range $M<M_{Fmin}$, and, then, they start to increase for $M>M_{Fmin}$. 

\begin{figure}[H]
\begin{center}
	\includegraphics[scale=0.5]{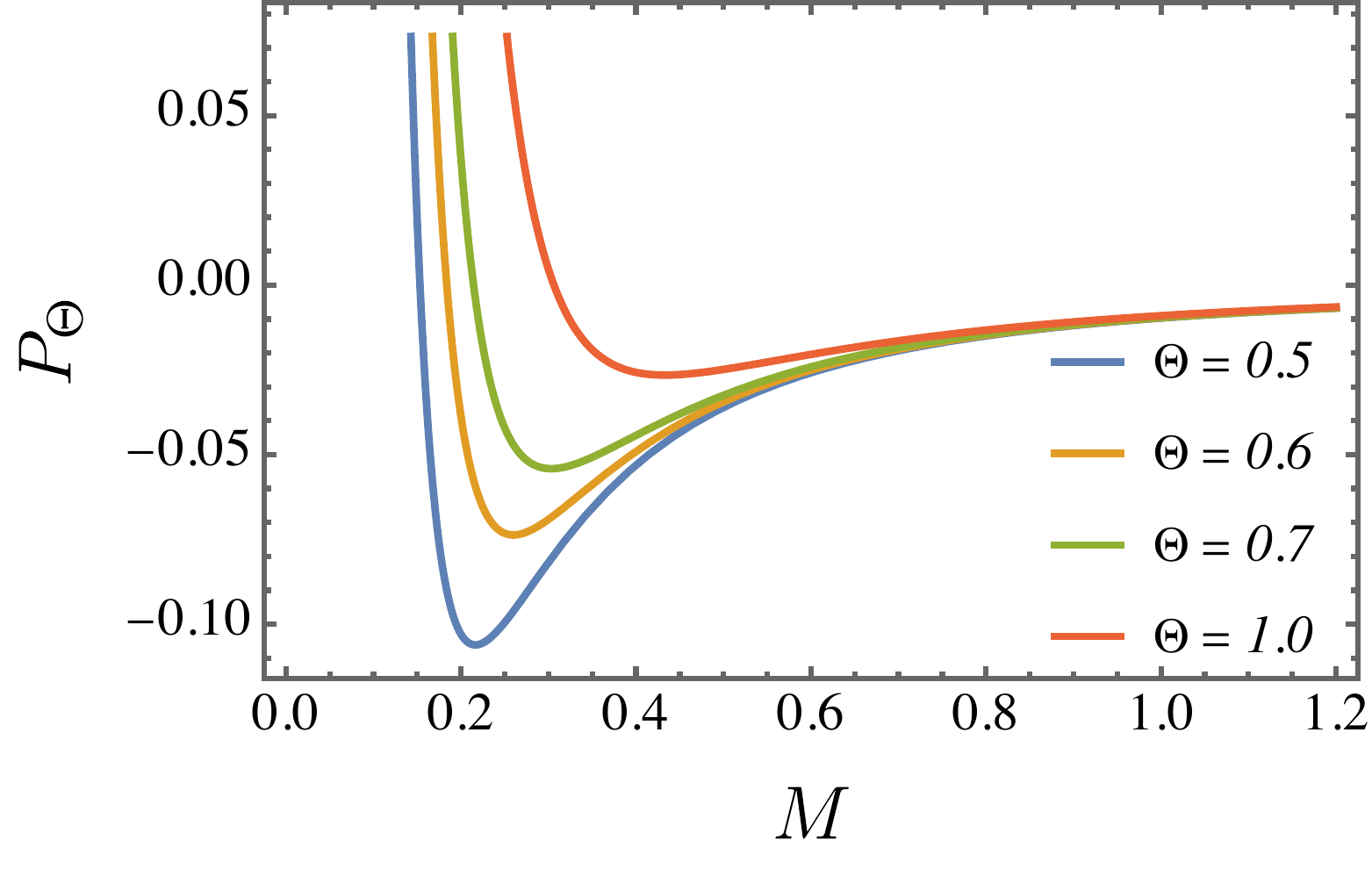}
 \includegraphics[scale=0.422]{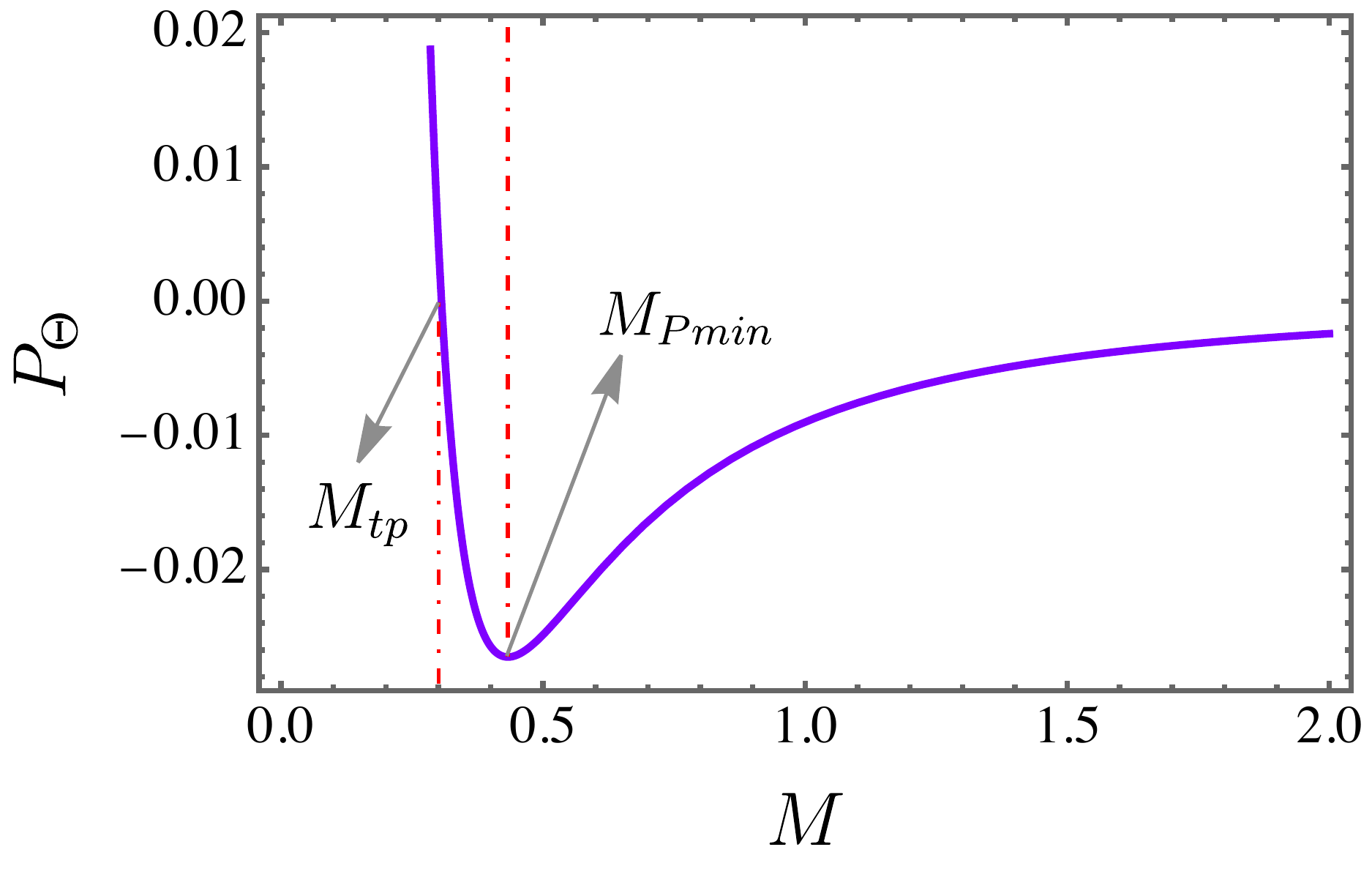}
	\caption{
		$P_{\Theta}$ as a function of the mass $M$ for different values of $\Theta$ (on the left). Also, it is also highlighted the behavior of pressure to $\Theta=1$ (on the right).
		\label{Fig5}}
\end{center}
\end{figure}

Now, let us apply a well--known standard expression to derive the pressure of the black hole:
\begin{equation}\label{DefPressure1}
P_{\Theta}=-\frac{\partial M_{\Theta}}{\partial V_{\Theta}} = -\frac{\partial M_{\Theta}/\partial r_{s\Theta}}{\partial V_{\Theta}/\partial r_{s\Theta}},
\end{equation}
which after substituting Eq. \eqref{DefFreeEnergy1} into Eq. \eqref{DefPressure1}, we get
\begin{equation}\label{DefpressureBH}
\begin{split}
P_{\Theta}&=  -\frac{1}{32\pi M^{2}}\left(1-\frac{3\Theta^{2}}{32M^{2}}\right).
\end{split}
\end{equation}
Note that this deformed pressure is also obtained  up to the second order of $\Theta$.
Referring to Figure \ref{Fig5}, it can be seen that for a differential value larger than the remnant mass, i.e., $ \delta M\equiv M_{rem}-M_{tp}$, where $M_{tp}$ is the transition point mass, being given by $M_{tp}=0.307$, the pressure value goes to zero. As a matter of fact, at $(0.307,0)$, there exists a physical transition point associated with the interaction of the corresponding black hole with the spacetime. This means that when the values of $P_{\Theta}$ are negative, the spacetime is under the pressure due to the black hole. On the other hand, when values of $P_{\Theta}$ are positive, the spacetime gives pressure to it. It is worth mentioning that this thermodynamic function decreases steeply until reaching its minimum value at $(0.433,-0.027)$. After that, no matter how much the mass increases, the values of $P_{\Theta}$ will not be positive.

Since the black hole begins to evaporate for either non--commutative or commutative cases, in order to obtain its ``lifetime'', we use the fact that the \textit{Stefan--Boltzmann} radiation law approximates to the rate of the energy loss as
\begin{equation}\label{EnergyLossRate}
\frac{\mathrm{d}M}{\mathrm{d}t} = -\tilde{\alpha} \sigma \tilde{a}  T^{4},
\end{equation}
where $\sigma$ being the cross--section area, $\tilde{\alpha}$ and $\tilde{a}$ are the greybody factor and the radiation constant respectively \cite{Ong2018JHEP}.
In particular, in the geometric optics approximation, it is known that the photon capture cross section is $\sigma = 27 \pi M^{2}$, in which $M=\frac{1}{8\pi T}$. Then, let us write 
\begin{equation}\label{EnergyLossRate1}
\frac{\mathrm{d}M_{\Theta}}{\mathrm{d}t} = -27\tilde{\alpha}\pi \tilde{a}M^{2}_{\Theta}T^{4}_{\Theta},
\end{equation}	
for our deformed black hole. 
In this framework, the mass evolution form becomes 
\begin{equation}\label{EnergyLossRate2}
\frac{\mathrm{d}M_{\Theta}}{\mathrm{d}t} = - \frac{\lambda}{M^{2}}\left(1-\frac{3\Theta^{2}}{64M^{2}}\right),
\end{equation}
with $\lambda=\frac{27\tilde{\alpha}\pi \tilde{a}}{\left(8\pi\right)^{2}}$.
Therefore, to estimate the time of the evaporation process, $t$, it is useful to express   
\begin{equation}
\int_{\tilde{M}}^{\tilde{M}_{i}}  M^{2}\left(1+\frac{3\theta^{2}}{64M^{2}}\right)       \mathrm{d}M = \lambda \int_{0}^{t} \mathrm{d}t,
\end{equation}
where $\Tilde{M}_{i}$ and $\Tilde{M}$ are the initial and the ``current"  mass after a time $t$ has passed respectively.
With above expression, we can solve $t$ yielding
\begin{equation}\label{evaporationtime1}
t=\frac{1}{\lambda}\left[\left(\frac{\tilde{M}_{i}^{3}}{3}+\frac{3\tilde{M}_{i}}{64}\Theta^{2}\right)-\left(\frac{\tilde{M}^{3}}{3}+\frac{3\tilde{M}}{64}\Theta^{2}\right)\right],
\end{equation}
which it clearly shows the dependence of time and masses. From this expression, as one can naturally expect, at any time after $t=0$, the mass of the back hole, $\tilde{M}$, will be smaller than its initial one. In this regard, if we consider the final state of the black hole (in a $t_{f}$ time) where its mass will be equal to $M_{rem}$, the black hole ``lifetime'' can be obtained in the following form
\begin{equation}
t_{f}= \frac{1}{\lambda}\left[\frac{M_{i}^{3}}{3}+\left(\frac{3M_{i}}{64}-\frac{\sqrt{3}}{128}\Theta\right)\Theta^{2}\right].
\end{equation}
Also, the time evaporation in the absence of the non--commutative effect is
\begin{equation}
\lim\limits_{\Theta\rightarrow 0}	t_{f} = \frac{M_{i}^{3}}{3\lambda}.
\end{equation}
Now, we can highlight a novel consequence of our results: an usual Schwarzschild black hole will have an finite ``lifetime'', i.e., the evaporation process terminates and part of the information will be lost \cite{NicoliniPLB2006,NozariCQG2008,PerezJMP2009}.

\section{Conclusion\label{Conc}}

Designing general relativity on a non--commutative spacetime, we presented the modified Einstein’s gravity arising from deformed tetrad ﬁelds by contracting the non--commutative gauge dS group, $\mathrm{SO}(4,1)$, with the Poincar\`{e} group, $\mathrm{ISO}(3,1)$, based on the SW map approach. According to the non--null components of the $\Theta^{\mu\nu}$ tensor, we expressed an exact spherically symmetric solution to the modified Einstein’s equation. The deformation of gravitational gauge potentials, achieved in $(3+1)$-dimensional non--commutative spacetime, established a deformed Schwarzschild black hole from the prominent modified metric tensor $\hat{g}_{\mu\nu} (x,\Theta)$.  
In this manner, the thermodynamic properties and the evaporation process in this context were investigated. We calculated the modified Hawking temperature $T_{\Theta}$ and the other deformed thermal state quantities: entropy, heat capacity, Helmholtz free energy and pressure. 

As it could be seen, the corrections emerged only considering the second order expansion of $\Theta$, which meant that no first--order terms affected our calculations.
Furthermore, when parameter ascribed to the non--commutativity vanished, namely, $\Theta \rightarrow 0$, we properly recovered the well--known thermodynamic properties of the standard Schwarzschild black hole.

Also, we investigated the evaporation process in this scenario, which stopped when the black hole was reduced to a stable Planck--size remnant mass. Such a feature was only possible due to the existence of the non--commutativity corrections. Therefore, the evaporation process discontinued and part of the information was maintained, being encapsulated by the remaining mass of our stable black hole. 





\section*{Acknowledgements}
The authors would like to thank the anonymous referee for a careful reading of the manuscript and for the remarkable suggestions given to us.
Most of the calculations were performed by using the \textit{Mathematica} software. Particularly, A. A. Araújo Filho is supported by Conselho Nacional de Desenvolvimento Cientíıfico e Tecnológico (CNPq) -- 200486/2022-5. P. J. Porfírio would like to thank CAPES for financial support. 


\section{Data Availability Statement}

Data Availability Statement: No Data associated in the manuscript


\end{document}